# Ge Microdisk with Lithographically-Tunable Strain using CMOS-Compatible Process

**David S. Sukhdeo[1], Jan Petykiewicz[1], Shashank Gupta[1], Daeik Kim[2], Sungdae Woo[2], Youngmin Kim[2], Jelena Vučković[1], Krishna C. Saraswat[1], and Donguk Nam[2,*]**

[1]*Department of Electrical Engineering, Stanford University, Stanford, CA 94305, USA*
[2] *Department of Electronic Engineering, Inha University, Incheon 402-751, South Korea*
[*]*dwnam@inha.ac.kr*

**Abstract:** We present germanium microdisk optical resonators under a large biaxial tensile strain using a CMOS-compatible fabrication process. Biaxial tensile strain of ~0.7% is achieved by means of a stress concentration technique that allows the strain level to be customized by carefully selecting certain lithographic dimensions. The partial strain relaxation at the edges of a patterned germanium microdisk is compensated by depositing compressively stressed silicon nitride layer. Two-dimensional Raman spectroscopy measurements along with finite-element method simulations confirm a relatively homogeneous strain distribution within the final microdisk structure. Photoluminescence results show clear optical resonances due to whispering gallery modes which are in good agreement with finite-difference time-domain optical simulations. Our bandgap-customizable microdisks present a new route towards an efficient germanium light source for on-chip optical interconnects.

## 1. Introduction

Germanium (Ge) has garnered much attention recently as a candidate material for optical interconnects [1–5] due to its inherent CMOS compatibility. Creating an efficiency Ge-on-Si light source, however, requires band engineering in the form of tin alloying [6–8] or tensile strain to overcome the limitations of Ge's indirect bandgap [9–11]. Initial research activities were focused on exerting large mechanical stress in Ge thin film membranes via water/gas pressures [12,13] and stressor layers [14], and >1% biaxial strain was successfully achieved. Also, there have been numerous recent reports on inducing large uniaxial tensile strain in Ge [15–18]. To create a Ge laser, the last missing piece for completing integrated optical interconnects, optical mirrors have also been integrated with highly strained Ge gain media [15,19–21]. Ghrib et al. first presented suspended Ge microdisk structures with stressed silicon nitride (SiN) that transfers large tensile strain to Ge [20]. However, such a structure suffers from strain inhomogeneity which causes several problems including gain broadening. More recently, similar structures but with spatially homogeneous strain distribution has been demonstrated by wrapping a stressed SiN layer all around Ge microdisks [21]. However, because the strain level is pre-determined by the thickness and residual strain of the bottom SiN layer, it is challenging to customize the strain level in each Ge microdisks. Since the ability to conveniently tune the strain level in Ge may allow the creation of multiple lasers operating at different wavelengths, the strain tunability can be one of the key functions for wavelength-division multiplexed (WDM) optical interconnects.

In this paper, we present a new approach towards a Ge optical resonator with lithographically tunable biaxial tensile strain. Strained microdisk can be created by patterning homogenously strained central region of the recently reported structure for which strain is concentrated due to geometrical effect [22]. The partial strain relaxation at the edges of a patterned Ge microdisk is compensated by depositing compressively stressed SiN layer, and we achieved homogeneous biaxial tensile strain of ~0.7% over the region where optical fields are strongly confined within Ge microdisks. Two-dimensional Raman spectroscopy measurements along with finite-element method (FEM) simulations confirm a relatively uniform strain distribution within the final microdisk structure. Photoluminescence (PL) results show clear optical resonances due to whispering gallery modes which are in good agreement with finite-difference time-domain (FDTD) optical simulations. Our bandgap-customizable microdisks present a new route towards an efficient Ge light source for WDM on-chip optical interconnects.

## 2. Device Fabrication

To create a Ge-based laser that generally requires larger optical or electrical pumping compared to compound semiconductor lasers [5,23], it is critical to achieve both strong optical confinement and efficient thermal dissipation. Strong optical confinement is normally associated with suspended geometries due the large refractive index contrast between Ge and air [15,24] while efficient thermal dissipation is normally associated with stiction-based Ge-on-Si geometries due to the large thermal conductivity of Si [25]. To meet these two requirements simultaneously, we must have the Ge layer ultimately in contact with a material that is reasonably thermally conductive, or at least more thermally conductive than air, and that also offers a large refractive index contrast to Ge. To this end, we have used a recently developed Ge-on-dual-insulators (GODI) substrate as shown in Fig. 1(a). The substrate has two dielectric layers, aluminum oxide ($Al_2O_3$) and silicon dioxide ($SiO_2$), sandwiched between Ge and the Si substrate in the starting material stack [19,26]. Whereas the thermally grown $SiO_2$ is ~800-nm thick to avoid tunneling of optical fields from Ge through $SiO_2$ into the Si substrate, the $Al_2O_3$ layer deposited by atomic layer deposition (ALD) is only 25-nm thick to facilitate the stiction of Ge to $SiO_2$ to eliminate an air gap for more efficient thermal dissipation.

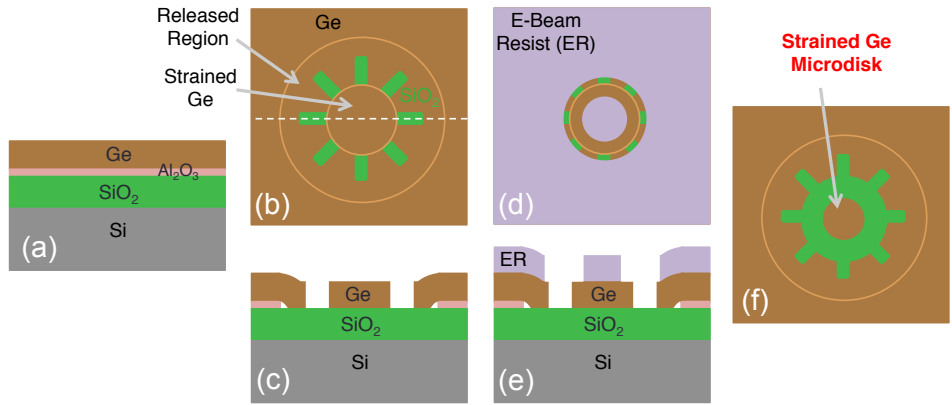

Fig. 1. Detailed fabrication process flow.

As shown in Fig. 1(b), we pattern a structure that can induce biaxial tensile strain within the central region of Ge upon undercutting the underlying layer [22]. The strain level can be fully customized by carefully selecting the ratio of inner diameter to outer diameter. Then, the underlying $Al_2O_3$ layer is isotropically and selectively removed in a wet potassium hydroxide (KOH) etch. This KOH etch leaves the $SiO_2$ layer intact so that in the final stiction step, Ge is now permanently adhered to the ~800-nm thick $SiO_2$ layer, resulting a final material stack of strained-Ge/$SiO_2$/Si as shown in Fig. 1(c). This now provides good optical confinement due to the large refractive index contrast between Ge and $SiO_2$ while also providing an efficient thermal dissipation path from Ge downward through $SiO_2$ into the underlying Si substrate.

Although SiO$_2$ may not be a particularly good thermal conductor, it is a dramatic improvement over a conventional air gap for optical confinement.

There remains the issue of the optical mode leaking away from the strained central region laterally through the connecting Ge layer. At first glance this may seem like a more fundamental challenge since the stress concentration approach requires the highly strained central region to be physically connected to the rest of the structure, as shown in Fig. 1(b), in order for the initial stress from the entire structure to be transferred to the central region [22]. However, after Ge is released and then permanently adhered to the underlying oxide by stiction, we find that the stress redistribution has already taken place and the Ge layer is now fixed in place by its bottom surface via van der Waals force. To increase the bonding force between Ge and oxide further, the sample is annealed at ~350°C. Using electron-beam lithography and dry etching, we pattern the strained central region into a more traditional circular resonator of ~3.5 µm in diameter as shown in Figs. 1(d), (e) and (f), that is laterally surrounded by air on all sides. This virtually eliminates any lateral leakage of the optical mode and enables an efficient optical resonator with whispering gallery modes (WGMs).

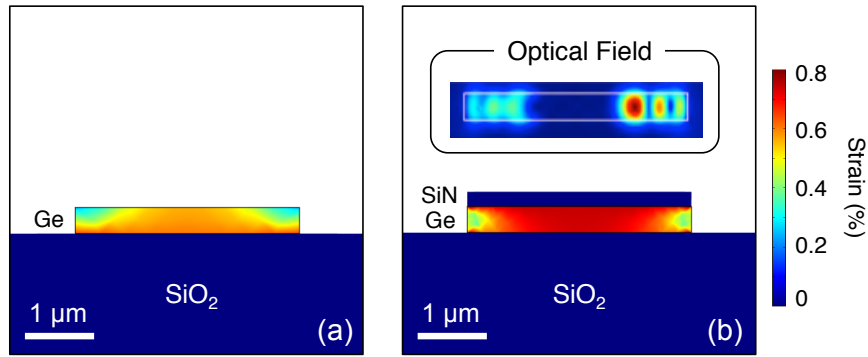

Fig. 2. (a) Simulated strain distribution of as-patterned strained Ge microdisk showing significant strain relaxation. (b) Simulated strain distribution of strained Ge microdisk with SiN stressor showing mostly restored strain homogeneity. Inset of Fig. 2(b) shows simulated optical field distribution of the structure in Fig. 2(b).

Upon removing the lateral connection to the rest of the Ge layer, the microdisk structure suffers from the strain relaxation at the top surface, particularly near the outer perimeter of the top surface as evidenced by FEM simulation shown in Fig. 2(a). This is common phenomenon in situations where the strain is maintained solely from the bottom surface. To compensate this strain relaxation, a compressively stressed SiN layer is deposited on Ge, resulting in a more homogenous strain distribution over a large area as shown in Fig. 2(b). The inset of Fig. 2(b) shows a simulated, cross-sectional optical field distribution of a higher-order WGM in our final structure. Although there still exists strain inhomogeneity especially near the edges of the microdisk, it is possible to achieve a good optical mode overlap with homogeneously strained region for especially higher-order WGMs as evidenced by comparing the strain and optical field distributions in Fig. 2(b).

## 3. Device Characterizations

Figures 3(a) and (b) show scanning electron micrographs of un-patterned and patterned structures with the stressed SiN layer on top, respectively. The lateral air gap between the microdisk and the remaining Ge layer is ~4 µm which is large enough to avoid any lateral optical tunneling. To confirm whether or not the large and spatially uniform biaxial strain can be achieved experimentally, 2D Raman mapping was performed. A 514-nm excitation laser was used and the power level was kept to minimum to avoid any possible heating effect. For the un-patterned structure, equivalent biaxial tensile strain of ~0.7% is induced over a relatively large area as shown in Fig. 4(a). For the patterned structure, however, the strain has become severely inhomogeneous as expected from the previous FEM simulation result. As shown in Fig. 4(b), while strain is partially relaxed to ~0.57% at the center, the outer edges are almost completely relaxed. Fig. 4(c) shows the Raman mapping of the final microdisk structure with the compressively stressed SiN layer on top. It is clear that SiN deposition can restore the desired strain homogeneity as also expected from the FEM simulation (Fig. 2(b)).

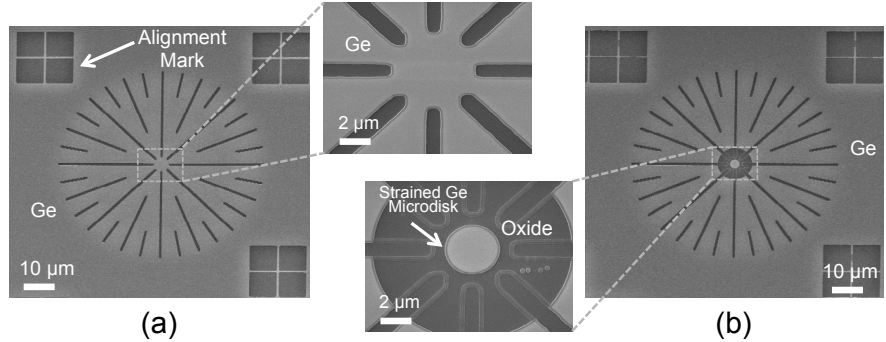

Fig. 3. (a) Scanning electron micrograph (SEM) of biaxially strained Ge structure before patterning. (b) SEM of biaxially strained Ge microdisk with SiN on top.

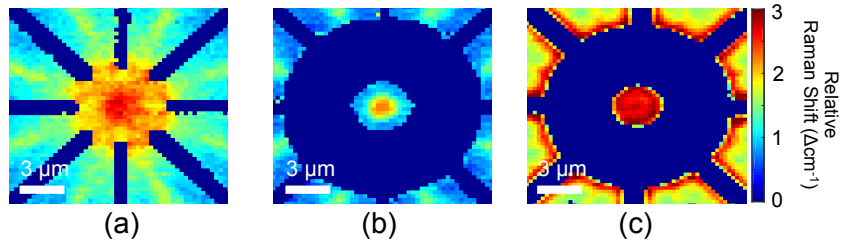

Fig. 4. Two-dimensional Raman mapping of (a) unpatterned structure, (b) as-patterned microdisk, and (c) final microdisk with SiN on top.

Having confirmed that our optical cavities can maintain the desired strain distribution, the next step is to demonstrate the presence of optical resonances in the light emission. We have done that by measuring the PL spectra from two microdisks, one with ~0.4% strain and another with ~0.7% strain near the center, before and after patterning our optical cavity in each case. As can be seen in Fig. 5(a), optical resonances are clearly visible when comparing PL spectra from patterned and un-patterned structures. The resonances are not particularly sharp, likely due to nitride surface roughness which may be improved with future optimizations, but establish a clear proof of concept that we have indeed integrated an optical cavity into our strained Ge layer while preserving the large, homogeneous biaxial strain. We have also performed FDTD optical simulations to prove that the optical resonances arise from WGMs within our microdisk. Figure 5(b) shows a simulated emission spectrum of our finalized structure with SiN on top. The resonance peak positions of the simulated spectrum are in reasonable agreement with the spectrum from the 0.4%-strained microdisk. The resonances from the 0.7%-strained microdisk are less clear and this can be possibly attributed to increased material absorption near the resonance wavelengths due to the reduced bandgap of Ge with larger tensile strain.

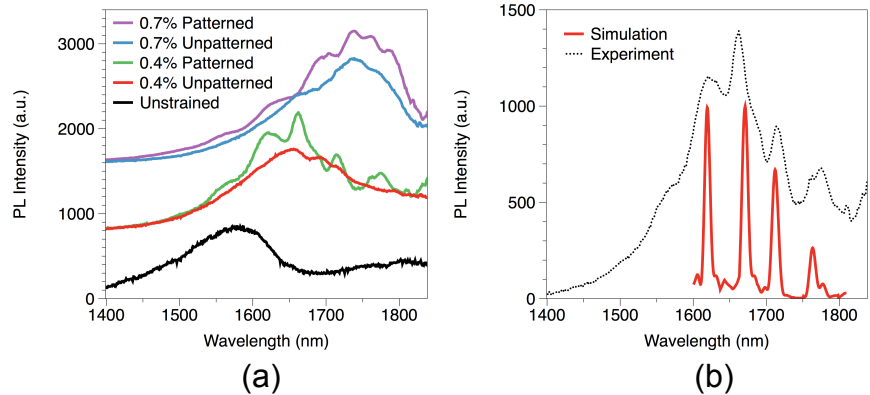

Fig. 5. (a) PL spectra from unpatterned and patterned structures for two different strain levels. (b) Simulated optical spectrum (red) and the spectrum from 0.4%-strained microdisk (black) for comparision.

**4. Summary**

We have demonstrated a CMOS-compatible approach to create strained Ge optical cavities, with biaxial strains of up to 0.7%. The large homogeneous biaxial strain was achieved following the approach of Ref [22], and then the optical cavity was realized by patterning our highly strained Ge into circular microdisk resonators. This circular microdisk patterning provides lateral confinement of the optical mode, with vertical confinement provided by a layer of $SiO_2$ between the Ge resonator and the underlying Si substrate. This $SiO_2$ layer also provides an efficient heat conduction path, ensuring efficient thermal dissipation which is critical for practical laser device applications. While patterning this cavity appears at first glance to be detrimental to the homogeneity of the strain distribution, we have shown that employing a final SiN stressor layer deposition can restore the spatial uniformity of the biaxial strain, resulting in a good optical mode overlap especially for higher-order WGMs. PL measurements along with FDTD optical simulations revealed that the optical resonances from strained Ge microdisks indeed arise from WGMs. Our strained Ge optical resonator structure represents a critical step toward realizing a highly strained Ge laser on Si for on-chip WDM optical interconnects.

**Acknowledgements**

This work was supported by the Office of Naval Research (grant N00421-03-9-0002) through APIC Corporation (Dr. Raj Dutt) and by a Stanford Graduate Fellowship. This work was also supported by an INHA UNIVERSITY Research Grant and by the Pioneer Research Center Program through the National Research Foundation of Korea funded by the Ministry of Science, ICT & Future Planning (2014M3C1A3052580).